# High power orbital-angular-momentum beam generation system based on coherent beam array combination technique

Dong Zhi,[1] Yanxing Ma,[1,2] Pu Zhou,[1,*] Rumao Tao,[1] Xiaolin Wang,[1] Lei Si,[1,3] and Zejin Liu[1]

[1]College of Optoelectronic Science and Engineering, National University of Defense Technology, Changsha, Hunan, P. R. China
[2]Email: xm_wisdom@163.com
[3]Email: w_zt@163.com
*Corresponding author: zhoupu203@163.com

## ABSTRACT

High power orbital-angular-momentum (OAM) beam has dominant advantages in capacity increasing and data receiving for free-space optical communication systems at long distance. Utilizing coherent combination of beam array technique and helical phase approximation by piston-phase-array, we have proposed a generation system of high power novel beam carrying OAM, which could overcome power limitation of common vortex phase modulator and single beam. We have experimentally implemented a high power OAM beam by coherent beam combination (CBC) of a six-element hexagonal fiber amplifier array. We show that the CBC technique utilized to control the piston phase differences among the array beams has a high accuracy with residual phase errors superior to $\lambda/30$. On the premise of CBC, we have obtained novel vortex beams carrying OAM of $\pm 1$ by applying an additional piston-phase-array modulation on the corresponding beam array. The experimental results approximately coincide with the theoretical analysis. This work could be beneficial to the areas that need high power OAM beams, like ultra-distance free-space optical communications, biomedical treatments, powerful trapping and manipulation under deep potential well.

## EXPERIMENTAL SETUP

The experiment platform based on CBC of fiber amplifiers (FA), as shown in Fig. 1, which can convert array beams into a vortex beam, is setup based on a six-element hexagonal fiber collimator array. As shown in Fig. 1, the seed laser for the fiber amplifiers is a linearly polarized, single frequency Yb-doped fiber seed laser (SL) with wavelength of 1064nm and power of 40mW. The seed laser amplified by a pre-amplifier (PA) and the output power is boosted to be 400mW. After that, a 1×8 fiber splitter (FS) has been used to split the pre-amplified laser into 8 sub-beams, from where we choose 6 sub-beams to carry out the following experiment. Before sending to the fiber amplifiers, each of the selected six fiber lasers is connected to a LiNbO$_3$-based fiber phase modulator (FPM). The gain fiber of the fiber amplifier is double clad fiber with core/inner cladding diameter of 10 μm/125 μm, based on which the power of each channel is amplified to 1W. Then a mode field adaptor (MFA, the input fiber is double clad fiber with core/inner cladding diameter of 10 μm/125 μm, and output fiber is double clad fiber with core/inner cladding diameter of 20 μm/400 μm) is used in each channel to connect the FPM with a fiber end-cap, which is spliced with 20 μm/400 μm delivery fiber and acts as the output terminal of fiber laser. Next, the array beams emitted from end-caps are collimated by a homemade collimator array with focal length of 800mm and clear aperture ($d$) of 58mm, which generate a collimated Gaussian beam array with the beam waist width $\omega_0$ to be 32.5mm and a beam truncation factor ($2\omega_0/d$) of 1.12. [1-3]

The collimated beam array then is separated into two parts: ① high power beam array with OAM and ② low power sampling beam array, when passing through a high reflective mirror (HRM1 in Fig. 1). The part ① is the principle generated high power vortex beam that is to be utilized in many fields, (e.g., long distance optical communication). While ② is the main phase-control part, which just needs a sample with small power proportion to generate a stable vortex-like stair phase array. Next, the sampling beam array ② is projected into a modified Newtonian reflector telescopic beam shrinking system in order to reduce the size of beam array smaller than the spatial light modulator (SLM), which is used here to precisely generate the required piston phase array and is controlled by a desktop computer. Just as depicted in Fig. 1, the shrinking beam array ③ is divided into two parts by using a beam splitter prism (BSP). The reflective part ④ is modulated and reflected by SLM. While, the transmission part ⑤ without applied piston phase array is focused directly by a focus lens (FL1) with focal length of 1m and is detected by a charge coupled device (CCD). If the impact of size is ignored, part ⑤ has the same phase and intensity distributions with part ①. Then part ⑥ is converted to be a beam array with a spiral staircase phase array. After reflected by HRM2 and focused by FL2 (focal length of 2m), the far-filed of part ⑥ is simultaneously detected by CCD2 and a photonic detector (PD), in front of which a pinhole with diameter of 80μm is attached.

The intensity distributions of two beams ⑤ and ⑥, which represent the situations before and after piston phase array addition and detected by CCD1 and CCD2 respectively, are observed by a laptop computer in real time. The PD with 150 MHz bandwidth in the experiment is used to provide feedback data in phase-locking control. The signal collected by PD is observed by an oscilloscope and processed by a FPGA controller. The output control signals produced from FPGA controller are applied on FPMs to control the array beams to an equiphase status. Throughout the experiment, single frequency dithering algorithm is used in the FPGA controller to perform the phase-locking control.

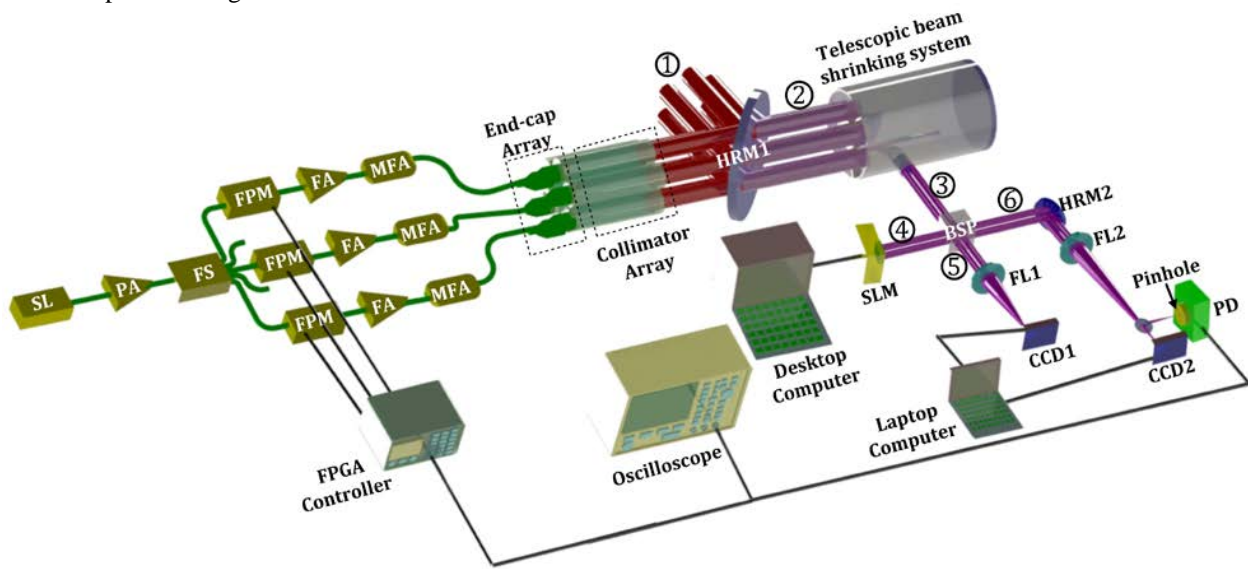

Fig. 1. Experimental setup of high power vortex beam generation system by array beams CBC technique. (SL: seed laser; PA: pre-amplifier; FS: fiber splitter; FPM: fiber phase modulator; FA: fiber amplifier; MFA: mode field adaptor; HRM: high reflective mirror; SLM: spatial light modulator; BSP: beam splitter prism; ①: high power beam array with OAM; ②: low power sampling beam array; ③: shrinking beam array; ④: beam array incident on SLM; ⑤: beam array without SLM phase modulation; ⑥: beam array applying SLM phase modulation; FL: focus lens; PD: photonic detector; CCD: charge coupled device; FPGA: field programmable gate array.)

## Points need to be mentioned

To the mentioned-above experimental setup, here are two points need to be explained:

The first one is the necessity of applications of fiber end-caps and MFAs. To the high power fiber laser, usually upon 1kW level, coreless end cap with anti-reflection film plated on the output end-face is usually employed as the output terminal of fiber laser [4-5]. The fiber end-cap can reduce the return electrical field intensity and avoid fiber facet damage, and then protect the forestage FA and SL. Due to that fiber with 20 μm core size has good power handling capacity (usually up to several kW [6-7]) than the fiber with small core diameter applied in the 1W-level FAs, double clad fiber with core/inner cladding diameter of 20 μm/400 μm is spliced on the end-cap to demonstrate a high power capability of the system.

As the mismatching of core diameters between the gain fiber in FA and the delivery fiber spliced on end-cap, the laser beam will inevitably excite a portion of high order modes when directly splicing the two fibers, which is not desirable in fiber laser CBC system. So in the following CBC experiments, we choose a fiber mode field adaptor (MFA) to connect the 10 μm /125 μm single mode fiber and the 20 μm /400 μm few-mode fiber, which could help to maintain the near fundamental mode operation [8].

The second one is the importance of beam shrinking system. As we know, the far-field diffraction angle of a collimated beam can be calculated by $\theta \propto \lambda/D$, where $\lambda$ is the wavelength and $D$ is the diameter of the collimated beam. Without the beam shrinking system, the size of the array beams must be very small in order to fit with the size of liquid crystal screen in SLM. As the liquid crystal screen is just a dozen millimeters, the size of incident array beams will be limited to millimeter level. Thus the diffraction angle of single beam will be about mrad level. That means when the receiver aperture is 1m, the propagation distance should be less than 1km. So this small array beams cannot be utilized in application scenarios that need long propagation distance. In our system, the size of collimated beam has been extended more than 30 times than that incident on SLM by the Newtonian telescope. Just as the experiment of OAM superposition modes of light transmitted through a 3 km intra-city channel in Vienna introduced in Ref. [9], they also use a telescope to expand the OAM beam to a diameter of approximately 60mm. And then project the bigger beam to free-space for propagation and reception. By careful calculations, the effective operating distance of our designed system reaches more than 60km with a 1m receiver. Thus this system has great potential in applications of optical communication system along a ground-to-space or space-to-ground path.

## EXPERIMENTAL RESULTS AND ANALYSIS

### Vortex beam generation

First, we experimentally test the availability of the real-time phase-locking (PL) control system based on single-frequency dithering technique [10-11]. Fig. 2 (a) shows the normalized voltage values detected by the PD in different situations to investigate the incoherent and coherent combining effects. We can see that without PL control, the voltage detected by PD fluctuates randomly along with time and its average value in more than 25 seconds is normalized to 0.346. When the active PL control is performing, the average value of the normalized voltage increases steeply up to above 0.95 in about 6.2 milliseconds. By carefully data analysis, we calculate the normalized mean voltage value is about 0.963 in above 25 seconds with root mean square (RMS) value $V_{rms}$ of 0.010. By using the equation of $\lambda \cdot (V_{rms})^{1/2}/\pi$, we can derive the residual phase errors to be about $\lambda/31$, which represents the PL control system is executing successfully and effectively. We also calculate the spectral density of power of phase noise with and without PL by Fourier transform of the voltage signal in time domain, just as shown in Fig. 2(b). From the clear comparisons of the two curves, we conclude that the phase noise with frequency below 100 Hz has been efficiently suppressed when the PL control is on.

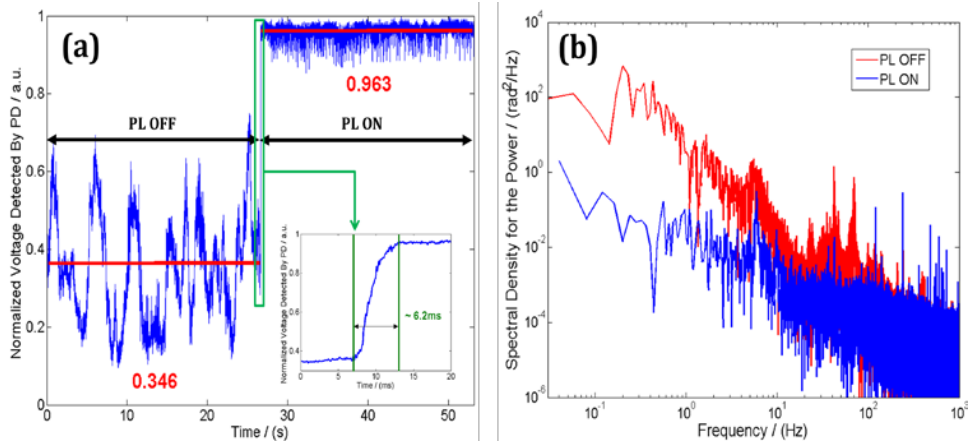

Fig. 2. The normalized voltage values detected by PD along with time and the corresponding spectral densities of power of phase noise. (a) Time-dependent normalized voltages. (b) Corresponding spectral densities.

With a strong foundation of PL control, which is the most important in CBC technique, we next carry out the vortex beam generation experiment through applying phase modulation (PM) by using SLM. The performances of the PL and PM both can be directly tested and observed by the long exposure beam pattern from the cameras CCD1 and CCD2. Figure 3 shows the intensity distributions in different situations just as the text descriptions in the left column, each one corresponds to a situation with or without PL and PM. The beam patterns in first column represent a 20 seconds long exposure detected by CCD2. As a monitor of PL, the patterns in CCD2 can only distinguish the incoherent and coherent states. That is to say, when PL is on, the beam array is in a steady coherent state no matter PM is on or off. As shown in the figure, the second and fourth columns respectively represent the two-dimensional and three-dimensional long exposure intensity distributions acquired by CCD1. The beam array detected by CCD1 truly reflects the high power beam array ① shown in Fig. 1. When PL and PM are both off, the phase differences among the array beams are time-varying and the long exposure beam pattern is close to a Gaussian beam, which is similar to the ideal incoherent combining situation shown in the first picture in the third column. When only the active PL is on, the array beams are controlled high-efficiently coherent with each other. The coherent pattern shows a significant increase of central-lobe intensity along with a high fringe contrast up to 95.4%. By calculations, the normalized maximum value of beam intensity distributions increases from 0.19 to 1, which means the beam brightness of CBC state increases almost 5.6 times comparing with the incoherent situation. From this point of view, the CBC experiment has reached 5.6/6=93.3% of the ideal calculation result. Further, when the PM is applied together with PL, the wavefront of the beam array is superposed an spiral-stair-like phase array on the basis of perfect coherence, and then the beam array carrying OAM is generated. Just as shown in the figure, the beam pattern has a non-circular hollow shape in the center with six petals around the central part. By careful data procession, the fringe contrast of the central hollow beam is about 73.5%. By observation and analysis, we can see that the far-field distributions of the OAM beam obtained in experiment basically match the ideal situation except for some small distinctions between the two shapes. The difference mainly comes from the high order mode induced by the MFA in each channel and the residual phase errors from PL control system. As the phase differences of the array beams are precisely controlled in a state that satisfies the required piston phase array, this hollow shape is stable in time domain.

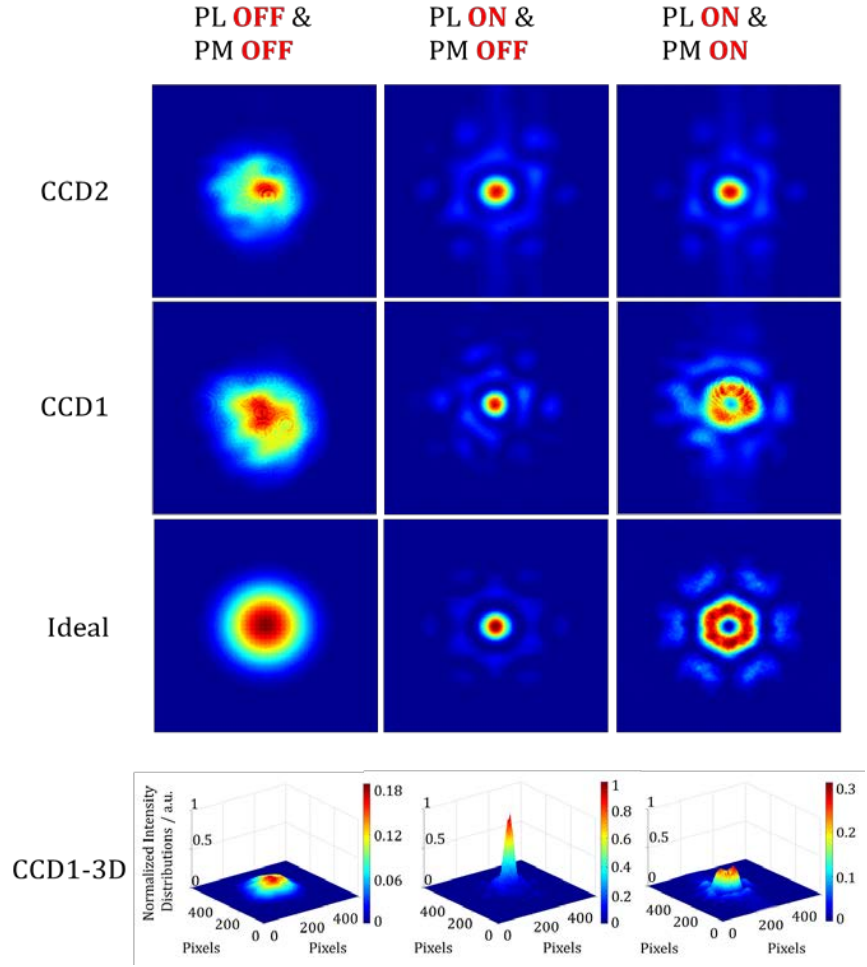

Fig. 3. The long exposure beam patterns in different situations.

## B. Spiral phase verification

To verify the phase of OAM beam obtained in the above experiment has a spiral distribution, we need to carry out another interference experiment to examine the OAM state. In general, the interference fringes for OAM verification have two types. One is the "Y"-shape interferogram generated from the interference of OAM beam and a plane wave beam. We can infer from the direction and numbers of the furcation in the interferogram that the beam carries how many OAMs. The other type is the helical structure, which comes from the interference between the vortex beam and a spherical wave. From observation of the direction and numbers of the furcation in this method, people can tell the OAM number of the measured beam.

In terms of the test of our generated vortex beam, the difficulty is to find a coherent laser beam with stable phase structure (plane or spherical). As the vortex beam is generated from an array beams, an arbitrary single beam with plane or spherical wavefront is not suitable for the interference experiment. As we know, beam ⑥ is an appropriate plane wave constructed by the same beam array under the CBC circumstance. Thus, to measure the OAM state in the vortex beam obtained in the previous section, we set up another experiment based on the original establishments. Just as depicted in Fig. 4(a), we separate a part of beam ⑥ using a beam splitter mirror (BSM) and HRM3. Then beams ⑤ and ⑥ are focused by FL1 together and the interference pattern at the focus point of FL1 is observed by CCD1.

When the OAM number $\ell=+1$, the beam array is applied a counterclockwise phase array, as shown in the red curve in Fig. 4(b). The corresponding theoretical and experimental interference patterns are depicted as Fig. 4 (c) and (d), respectively. The special interferogram has a fork fringe with an upside-down 'Y' shape in the central, which represents the number of OAM carried by the beam array is +1. The position of bifurcation is just the location of phase singularity. When we apply a clockwise phase array on the beam array, as shown in Fig. 4(e), the central part of the interferogram converts to an erected 'Y' shape fringe (Fig. 4(f) and (g)). Through the testing experiment, we can get that the generated vortex beam carries the OAM as we imagined, and the topological charge is consistent with that imposed by the SLM.

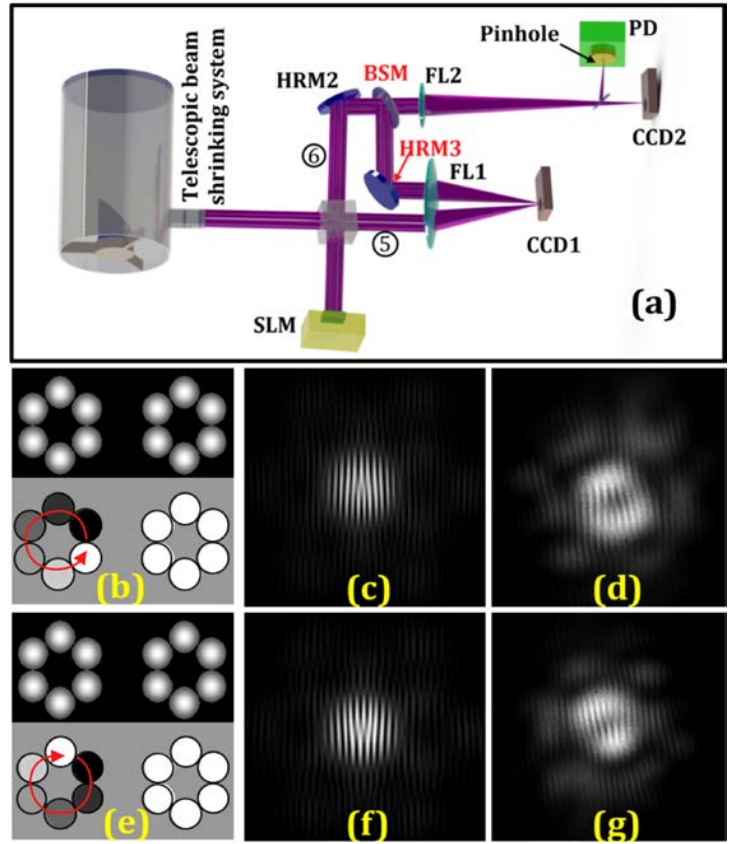

Fig. 4. The experimental setup (a) and interference results (b~g) of the OAM measurement. (b) is the intensity and phase distributions of beams ⑤ and ⑥ with the number of OAM to be $\ell=+1$; (c~d) are the corresponding interference patterns of theoretical and experimental results. (e~g) are the results refer to the situation of $\ell=-1$.

## CONCLUSIONS

In this paper, we have proposed a generation method of high power vortex beam using array beams coherent combination technique. We have experimentally implemented a high power beam carrying OAM by CBC of a fiber amplifier array. The CBC technique utilized to control the piston phase differences among the array beams has a high accuracy with residual phase errors of about λ/31. Based on CBC and desired piston phase applying by SLM, we finally obtain novel vortex beams carrying OAM of ±1, both basically coincide with the ideal situations.

As the array-beams-generated OAM beams with different eigenstates are orthogonal with each other, they can be well used to extend the signal channels in optical communication systems. The high power OAM beam can be used in many applications besides free space communications, such as high resolution imaging, biomedical engineering, special material processing, optical micromanipulation and so forth. Especially for the destruction of necrotic biological tissues in fields of biomedical treatment, the high power vortex beam (like ring Airy-Gaussian beam) has great potential and advantages [12]. Additionally, the microparticales under the radiation of high power OAM beam may show some attractive performances.

## ACKNOWLEDGMENT

The authors would like to acknowledge the support of the programs for the National Science Foundation of China under grant No. 61405255, No. 61378034 and No. 11504424; Graduate Student Innovation Foundation through the National University of Defense Technology, Changsha, China (B150705).

We also thank our colleagues Ph.D candidates Lei Li and Kun Xie, Dr. Wuming Wu, Dr. Zilun Chen and Dr. Yu Ning for their help with the experiment.